\documentclass[slac_one]{revtex4}
\usepackage{graphicx}
\usepackage{fancyhdr}
\begin{document}

\title{Hard Probes of the ATLAS Heavy Ion Physics Program} 

%

\author{Jiangyong Jia (for the ATLAS Collaboration)}
\affiliation{Chemistry Department, Stony Brook University, Stony
Brook, NY 11794, USA} \affiliation{Physics Department, Brookhaven
National Laboratory, Upton, NY 11793, USA}

\begin{abstract}
We review the physics performance of the ATLAS detector for hard
probes in heavy ion collisions, focusing on three topics: jet
reconstruction, direct photon measurement, and heavy flavor jet
tagging.
\end{abstract}

\maketitle

\thispagestyle{fancy}


\section{Introduction} 
The success of the RHIC program has been based largely on its
ability to deduce the properties of the dense medium using various
energetic ``hard probes'', including light quark jets
($u$,$d$,$s$), gluon jets, heavy quark jets ($c$,$b$) and photons
($\gamma$), which interact with different strength in the medium.
The interaction of the colored probes, i.e. quarks and gluons, not
only leads to the suppression of the probes themselves, it also
seems to initiate a medium response that is tightly correlated with
original jet direction, e.g. the ``ridge'' and the ``double hump''
phenomena observed in correlation analyses~\cite{Adare:2008cq}.
These experimental findings have inspired many theoretical
approaches to understand the medium response to hard probes. These
models include not only the more conventional tools such as
hydrodynamical and cascade models, with input from lattice QCD and
perturbative QCD energy loss~\cite{Muller:2006ee}, but also more
recent developments based on the AdS/CFT
duality~\cite{Herzog:2006gh,Gubser:2006bz}.

However, existing measurements of hard probes at RHIC have not
provided sufficient quantitative constraints on the detailed
properties of the dense medium. Most measurements are done with
leading particle spectra or two particle correlation analyses,
rather than a full jet reconstruction.  They thus reflect a
convolution of parton distribution functions, energy loss and
fragmentation, such that detailed information on energy loss
processes can not be extracted cleanly. Furthermore, the mechanisms
of medium responses are not well understood, in part due to
experimental uncertainties associated with the signal extraction,
and primarily due to the lack of clear theoretical constraints from
these observables. Finally, other hard probes, such as $\gamma$-jet
correlations, and heavy quark energy loss are limited by the low
rate, large background and the lack of direct jet reconstruction.

The LHC, with its center of mass energy 27 times that of RHIC,
greatly increases the kinematic reach and the rates for jets, heavy
flavor jets, and direct photons. These hard probes can be measured
precisely using the large-acceptance, multi-purpose ATLAS detector.
Recently, extensive simulation studies have been performed,
focusing on the following four areas as part of the ATLAS heavy ion
physics program:
\begin{enumerate}
\item ``Day-1''  measurements to probe bulk properties:
    multiplicity, collective flow, inclusive particle spectra;

\item Jet and photon measurements to probe jet quenching and
    medium response: full jet reconstruction, fragmentation
    functions and jet profile, di-jet correlations, direct
    $\gamma$, $\gamma$-jet correlations and heavy flavor jets;

\item Upsilon and $J/\Psi$ to probe Debye screening;

\item Low-$x$ physics at forward $\eta$ to probe the initial
    condition: Jet, particle spectra, and particle and jet
    correlations at forward $\eta$.
\end{enumerate}

Of these four topics, jet measurements are a particular strength of
the ATLAS heavy ion program. The ATLAS detector~\cite{atlas} has
been demonstrated to perform excellent measurements of jets,
photons, and muons even in a high-multiplicity heavy ion
environment. The ATLAS detector has multiple layers of
electromagnetic and hadronic calorimeters covering $2\pi$ in
azimuth and $-5<\eta<5$. The segmentation of the calorimeters
allows full jet reconstruction up to $|\eta|<5$, and direct photon
measurement up to $|\eta|<2.5$. Charged hadrons can be measured in
$|\eta|<2.5$ with high efficiency, which can be used to construct
jet fragmentation functions. By combining reconstructed tracks with
muons, it is possible to tag heavy flavor jets in $|\eta|<2.5$.
This paper will thus focus on the capabilities of ATLAS in jet
physics.

\section{Jet Physics}~\label{sec:jet}
The standard jet reconstruction algorithms developed for $e^+e^-$
and $p+p$ collisions need to be modified to account for the large
underlying background in Pb+Pb collisions. This background needs to
be properly estimated and subtracted before jet reconstruction, and
any fake jets due to background fluctuations also need to be
rejected. The following steps are used to evaluate the jet
performance in Pb+Pb collisions. A sample of PYTHIA di-jet events
is embedded into a sample of central HIJING Pb+Pb events, which are
used to model the expected underlying background. A $Q^2 <100$
GeV$^2$ cut is required for the HIJING Pb+Pb events to suppress
hard-scattering processes. The merged events are then run through a
GEANT4 simulation of the ATLAS detector and digitized into raw data
format, which is subsequently analyzed with the software used for
real data. An $\eta$ and $\phi$ dependent average underlying
background is calculated from the merged event, excluding cones
around jet seeds, and subtracted from the calorimeter cells.
Finally, a seeded-cone jet algorithm with seed energy of 5 GeV is
run on the background subtracted event to reconstruct jets. It is
found that the jet energy resolution in central Pb+Pb events
($dN_{ch}/d\eta$=2700 at mid-rapidity) is 25\% at 50 GeV, 15\% at
100 GeV, and saturates at about 10\% above 200 GeV. This can be
compared with $p+p$ collisions, for which the jet energy resolution
is about 10\% at 100 GeV and saturates to 7\% at large $p_T$.
Interestingly, the energy resolution was found to be almost
independent on the pseudo-rapidity in $|\eta|<5$, thanks to the
decrease in the background level in forward $\eta$ which almost
compensates the worsening of the calorimeter segmentation. The
position resolution of the jet is always better than 0.045 radians
at $E_T>50$ GeV even for the most central events.

\begin{figure}[ht]
\begin{tabular}{lr}
\begin{minipage}{0.68\linewidth}
\centering
\hspace*{-0.3cm}
\includegraphics[width=125mm]{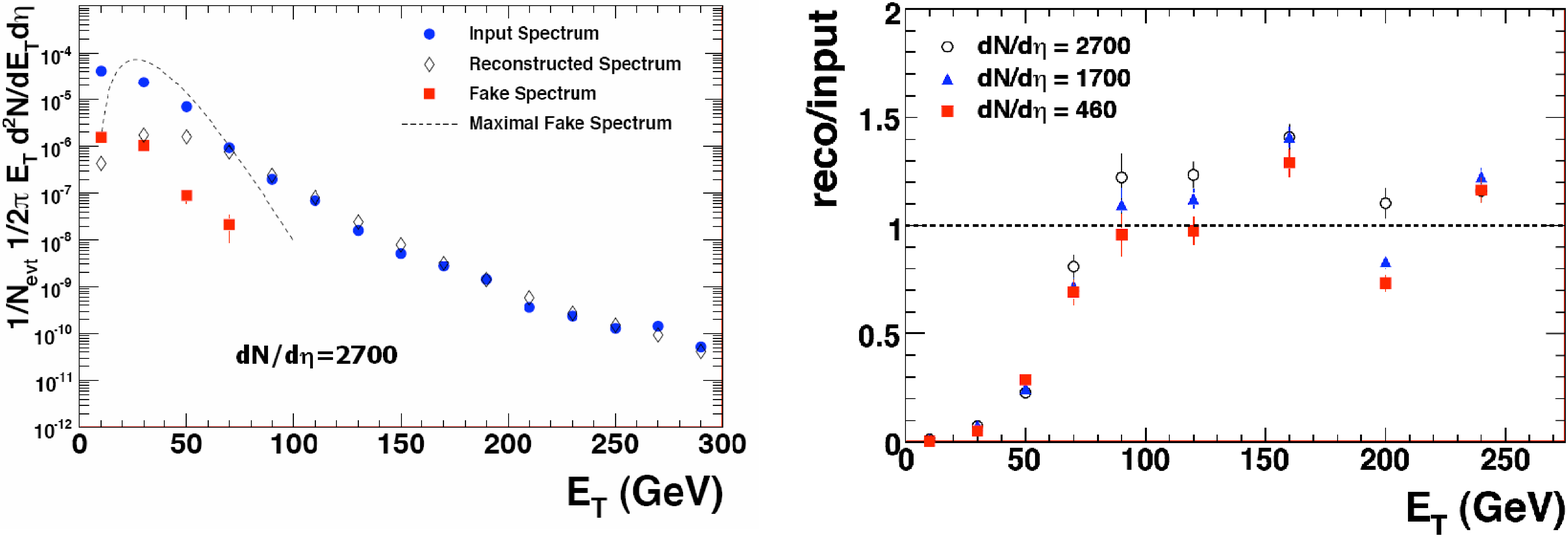}
\end{minipage}
&
\begin{minipage}{0.32\linewidth}
\vspace*{-0.5cm}
\caption{\label{fig:1} (left) Input (circles), reconstructed (diamonds), and fake (squares) spectra
for cone jets in central ($dN_{ch}/d\eta$ = 2700) Pb+Pb collisions. (right)
Ratio of reconstructed to input jet spectrum for three different Pb+Pb collision centralities without
efficiency and resolution corrections (the non-statistical fluctuations are due to procedures
in combining datasets generated for different $E_T$ ranges).}
\end{minipage}
\end{tabular}
\end{figure}

Fake jets are rejected by cutting on $\langle\sin R\rangle = (\sum
E_{T,cell} \sin R_{cell})/E_{T,jet}$~\cite{Grau:2008ed}, where the
$R_{cell}$ is the 2-D angle between the cell and jet axis.
$\langle\sin R\rangle$ approximates the width of the jet cell
energy, and it shrinks with jet energy for real jets but remains
roughly constant for fake jets. Fig.\ref{fig:1}(left) shows the
reconstructed jet yields in central Pb+Pb events, compared to the
input distribution. The efficiency, which is around 50\% around 70
GeV and saturates to 100\% above 100 GeV, is governed mainly by the
seed threshold (5 GeV) and the cut used to reject fake jets.

\begin{figure}[ht]
\begin{tabular}{lr}
\begin{minipage}{0.8\linewidth}
\centering
\hspace*{-0.3cm}
\includegraphics[width=145mm]{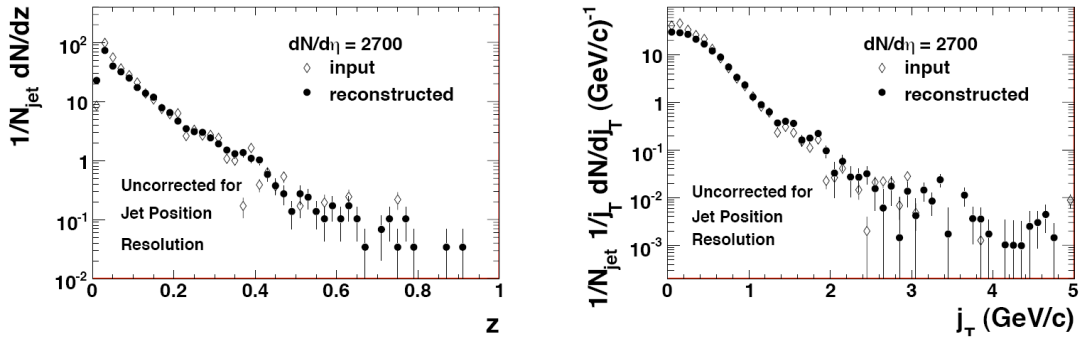}
\end{minipage}
&
\begin{minipage}{0.2\linewidth}
\caption{\label{fig:2} Input (open) and reconstructed (closed) fragmentation functions (left panel) and $j_T$ distributions (right panel). The reconstructed
quantities are before position and energy resolution.}
\end{minipage}
\end{tabular}
\end{figure}

The modification of jet properties is quantified by comparing jet
rates in Pb+Pb events to those from $p+p$.  Thus the level to which
one can recover an unmodified jet in heavy ion collisions is an
important indicator for the ATLAS sensitivity to jet modifications.
Fig.~\ref{fig:1}(right) suggests that the jet cross section can be
recovered at the 20\% level. Fig.~\ref{fig:2} shows the performance
of jet algorithm in measuring jet fragmentation functions and $j_T$
distributions~\cite{Grau:2008ef}. Both are important physical
observables for studying jet modifications in-medium. The former
characterizes the jet multiplicity, while the latter characterizes
the jet shape. Fig.~\ref{fig:2} shows that these distributions are
recovered within 20\% relative to those for the input jets. Thus
ATLAS should be sensitive to any medium effects that lead to more
than 20\% modifications.

One advantage of full jet reconstruction is that once the jet is
reconstructed, the jet direction and energy can be correlated with
the associated hadrons in the jet cone, which should provide direct
information on the jet medium response. By correlating the jet with
soft hadrons in the same hemisphere, one should be sensitive to the
near-side ``ridge'' phenomenon. By correlating the leading jet on
one side with those soft hadrons on the away-side, one should be
sensitive to the ``double hump'' observed at RHIC.

\section{Photon Physics}
The ATLAS detector also has excellent capabilities for direct
photon and direct photon-jet correlation measurements. The first
layer of the ATLAS barrel EM calorimeter is extremely finely
segmented in the beam direction, with a width of only 3mrad. A
single photon normally leaves behind a single peak about 3 strips
across, while a $\pi^0$ or $\eta$ meson normally fires multiple
strips with multiple peaks. To give an intuitive feeling, the strip
width, 3mrad, is the minimum opening angle for 50 GeV $\pi^0$. So a
large fraction of $\pi^0$ and most of $\eta$ mesons below 200 GeV
can be rejected based on their shower shapes.

We also reject decay photons using isolation cuts. Direct photons
are relatively isolated, while decay photons are normally
associated with the other high-$p_T$ hadrons coming from the same
original jet. A set of isolation criteria is used to reject decay
photons while preserving most of the direct photons. By optimizing
the cuts on the momentum of nearby charged particles and the level
of calorimeter energy flow in a cone of 0.2 radians around the
photon direction, one can achieve a relative rejection factor of
10-20 with an efficiency of 50-60\% for direct photons.

\begin{figure}[ht]
\begin{tabular}{lr}
\begin{minipage}{0.65\linewidth}
\centering
\includegraphics[width=115mm]{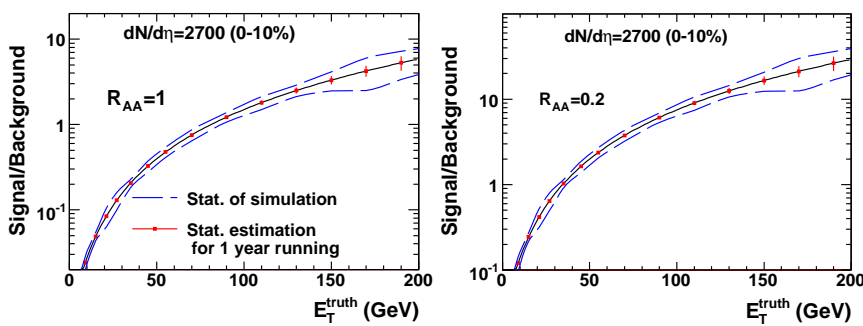}
\end{minipage}
&
\begin{minipage}{0.35\linewidth}
\vspace*{-0.3cm}
\caption{\label{fig:3} Direct photon signal to background (S/B) vs. $E_T$ for neutral hadron $R_{AA}$ = 1.0 (left)
and 0.2 (right) in 0-10\% central Pb+Pb collisions. The error bars indicate the expected statistical error including errors
in the background measurement and subtraction calculated for one year nominal run (0.5 nb$^{-1}$); the
dashed lines indicate the size of the statistical errors in our current simulation.}
\end{minipage}
\end{tabular}
\end{figure}

The total signal/background (S/B) ratio for the combination of the
shower shape cut and the isolation cut is shown in
Fig.~\ref{fig:3}. The S/B reaches 1 at around 100 GeV. However if
one accounts for the observed suppression of hadrons by applying a
suppression factor is 5 to the hadron yields in central Pb+Pb
collisions, then one reaches S/B=1 already at around 30 GeV. Even
at 200 GeV the statistical error expected for one month of nominal
Pb+Pb running per year is only 30\%, so one should be able to
measure the direct photon spectra out to 200 GeV.

The large yield of clean direct $\gamma$ can be used for
$\gamma$-jet correlation studies. A first look at the $\gamma$-jet
correlation has been done in Ref.~\cite{Grau:2008ef}. Since the
direct photons do not suffer geometrical bias, such correlation
studies allow a quantitative study of the path length dependence of
the jet energy loss and the medium response.  They can also help to
extract the jet fragmentation function at low $z$, as well as help
to tune the jet reconstruction algorithms at low $E_T$ ($<50$ GeV).

\section{Heavy Quark Jets}
In ATLAS, heavy quark jets can be tagged using the displaced vertex
measured by the silicon trackers. A second, complementary method is
to tag heavy quark jets by bottom and charm hadrons via their
semi-leptonic decay channels: $c,b\rightarrow D,B \rightarrow \mu +
X$. ATLAS can measure single muons above 3 GeV and in
$|\eta|<2.5$~\cite{sasha}. This allows tagging of jets originated
from heavy quarks. To leading order in QCD, most of the muon-tagged
jets come from hard-scattering processes that created a $c\bar{c}$
or $b\bar{b}$ pair. However a fraction of the muons may come from
the weak decay of light hadrons and are not associated with heavy
quark jets. The performance of muon tagging for heavy quark jets
can be quantified by the purity of heavy quark jets in the tagged
sample, and the tagging efficiency for jets that are known to come
from heavy quarks.

\begin{figure}[ht]
\begin{tabular}{lr}
\begin{minipage}{0.75\linewidth}
\centering
\includegraphics[width=125mm]{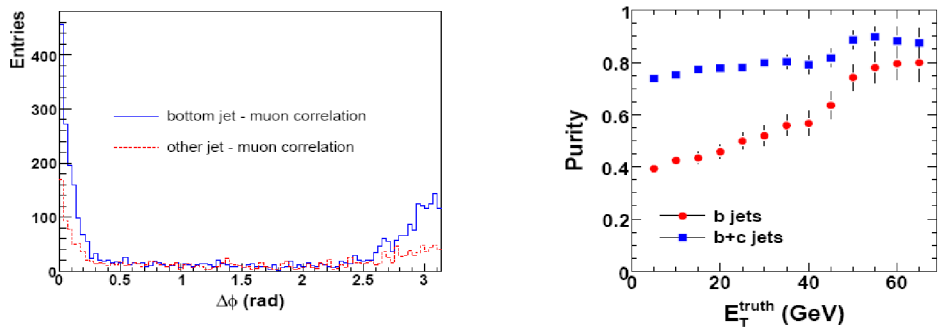}
\end{minipage}
&
\begin{minipage}{0.25\linewidth}
\caption{\label{fig:4} (left) Azimuthal correlation between the reconstructed jets and the muons.
(right) Bottom jet and heavy quark purity requiring a muon as a function of truth muon $E_T$. }
\end{minipage}
\end{tabular}
\end{figure}

To estimate the purity of heavy quark jets in the tagged jet
sample, PYTHIA minimum bias events were generated with the
requirement that each event contain at least one muon with $p_T >
5$ GeV/c and one jet with $E_T > 35$ GeV. The resulting events were
then embedded into central HIJING Pb+Pb events ($dN_{ch}/d\eta$ =
2700 at mid-rapidity) and the jets were reconstructed as described
in Sec.~\ref{sec:jet}. Single muon candidates were reconstructed
using the standard tracking and muon identification software in
ATLAS. To identify the heavy quark jets, the jets reconstructed
from the merged event are matched to the truth jets from input
PYTHIA events by requiring their opening angle to be $\Delta \phi<
0.2$ rad. Truth jets are tagged as bottom, charm or light quark
jets by tracing the PYTHIA ancestry information back to the
original string. Fig.~\ref{fig:4}a shows the azimuthal correlation
between the tagged jets and the muon. A cut of 0.16 rad gives a
70\% tagging efficiency while relaxing the cut to 0.32 rad gives an
efficiency of 80\%. Fig.~\ref{fig:4}b shows the purities of the
muon-tagged jets as a function of the trigger muon $p_T$. For muons
at $p_T > 60$ GeV/c, approximately 80\% of all tagged jets are
bottom-jets. Clearly, the purity of the heavy quark jets can be
enhanced by cutting on the muon-jet $\Delta\phi$ and the muon
$p_T$.

\section{Summary}
In summary, the ATLAS detector provides unprecedented capability
for probing the dense medium via jets, photons, and heavy quarks.
Detailed simulations have been carried out to estimate the physics
performance for jets and direct photons.  It is shown that jets can
be reconstructed above 50 GeV with reasonable resolution, low fake
rate and high efficiency. Jet spectra, jet fragmentation functions
and jet shapes can be recovered within 20\% of the input
distribution before applying any corrections. Using the highly
segmented first layer of the ATLAS calorimeter, combined with
isolation cuts, we show that direct photons can be measured with a
good S/B ratio up to 200 GeV for a nominal LHC year. By tagging on
high $p_T$ single muons, one can obtain a sample of high $p_T$
heavy quark jets. These capabilities are crucial for understanding
the energy loss and medium response mechanisms, and should provide
quantitative constraint on the properties of the hot and dense QCD
matter.

The large acceptance ATLAS detector also provides precision
measurements of global event characteristics, such as impact
parameter, number of participant, number of collisions and reaction
plane. These global observables can be measured independently using
various subsystems~\cite{ps:2008}, which help to minimize the
systematic errors. They shall open up the possibility of a detailed
study of the dependence of energy loss and medium response on
collision geometry.

\end{document}